\documentclass{article}
\usepackage{spconf,amsmath}
\usepackage{graphicx}
\usepackage{float}
\usepackage{amsfonts}
\usepackage{amsmath}
\usepackage{amssymb}
\usepackage{cite}
\usepackage{bm}
\usepackage{mathrsfs}
\usepackage{epsfig}
\usepackage{algorithm}
\usepackage{algorithmic}
\usepackage{amsthm}
\usepackage{color}
\usepackage{caption}
\usepackage{subcaption}
\usepackage{subfig}

\ninept
\begin{document}

% Title.
% ------
\title{Detection with  Multimodal Dependent  Data Using  Low Dimensional Random Projections}
\name{Thakshila Wimalajeewa and Pramod K. Varshney}
\address{Syracuse University, Syracuse, NY USA}
%\author{\authorblockA{Thakshila Wimalajeewa  \emph{Member, IEEE}, and Pramod K.
%Varshney, \emph{Fellow IEEE}}
%}

\maketitle\thispagestyle{empty}

\begin{abstract}
 Performing likelihood ratio based detection with high dimensional   multimodal  data is a challenging problem  since the  computation of  the joint probability density  functions (pdfs)  in the  presence of inter-modal  dependence is difficult. While some  computationally expensive approaches have been proposed for dependent  multimodal data fusion (e.g., based on copula theory), a commonly   used tractable approach is to  compute the joint pdf  as the product of marginal pdfs  ignoring  dependence. However, this method leads to poor  performance when the data is strongly  dependent.  In this paper, we consider the problem of detection  when  dependence among multimodal data is modeled in a compressed domain where  compression is obtained   using  low dimensional  random projections. We employ a Gaussian approximation while modeling inter-modal dependence in the compressed domain which is computationally more efficient.  We show that, under certain conditions, detection with  multimodal dependent  data  in the compressed domain with a small number of compressed measurements  yields  enhanced performance compared to detection with   high dimensional data  via either  the product approach or  other suboptimal fusion  approaches proposed in the literature.
 %We futher briefly discuss how to determine the
 %Thus, the advantageous of the proposed approach are two fold:   (i). it has  less computational complexity since fusion is performed in the  compressed domain, and (ii). it is capable of providing enhanced performance compared to  the product approach with uncompressed data.
\end{abstract}
{\bf Index terms}: Compressive sensing, multimodal data, inter-modal dependence,  likelihood ratio based detection, copula theory

\section{Introduction}
Fusion of high dimensional heterogenous data for different inference problems is challenging  in many applications \cite{Lahat_Proc2015}. While likelihood ratio (LR) based detection (with no unknown parameters) is optimal in the  Bayesian setting, its optimality is not guaranteed when  the exact joint probability density function (pdf)  is not available. It is difficult to compute the joint pdf  in the presence of multimodal dependence  unless data can be modeled as
Gaussian. To model complex dependencies among multivariate data in order to compute the joint pdf,  copula theory has been used in \cite{Mercier_2007,iyengar_tsp11,ashok_tsp11,ashok_taes11,Iyengar2011,Subramanian_2011,He_tsp2015}. While there are several copula density functions  available in the literature, finding the best copula function that fits a given set of data is  computationally challenging.
%This is because different copula functions may characterize different types
%of dependence behaviors  among  random variables \cite{Nelsen2006,Mari_B1}.
Further, in order to fuse multimodal data with more than two modalities, finding multivariate copula density functions is another challenge since most of the existing copula functions are derived considering the bivariate case. Thus, the benefits of the use of copula theory for LR  based detection with  multimodal data comes at a higher computational price. One of the commonly used suboptimal methods  for  fusion of multimodal data is to neglect inter-modal dependence and   compute the likelihood ratio only  based on  the marginal pdfs   of each modality (we call this 'the product approach' in the rest of the paper). However, this approach is expected to lead to poor performance when  inter-modal dependence  is strong.

To overcome the computational difficulties in  the fusion of  high dimensional  multimodal data for detection, in this paper, we consider the fusion problem  in a compressed domain where compression is achieved via low dimensional random projections as proposed in the compressive sensing (CS) literature \cite{candes1,candes2,donoho1,Eldar_B1}. The problem of detection with compressive measurements has been addressed by several recent works \cite{duarte_ICASSP06,
haupt_ICASSP07,davenport_JSTSP10,Wimalajeewa_asilomar10,
Gang_globalsip14,Bhavya_cscps14,Bhavya_asilomar14, Rao_icassp2012,Cao_Info2014,Kailkhura_WCL16,Kailkhura_TSP16,Wimalajeewa_tsipn16}.
 While some of the work, such as \cite{duarte_ICASSP06,haupt_ICASSP07,Gang_globalsip14,Rao_icassp2012,Cao_Info2014,Wimalajeewa_tsipn16} focused  on sparse signal detection,  some other works \cite{davenport_JSTSP10,Wimalajeewa_asilomar10,Bhavya_cscps14,Bhavya_asilomar14} considered  the problem of detecting  signals which are not necessarily sparse. When the signals are not necessarily sparse, it was observed that there is a performance loss when performing LR  based detection  in the compressed domain compared  to that with uncompressed data. However, when the signal-to-noise ratio (SNR) is sufficiently large, this loss is not significant and the compressed detector,  i. e., the detector based on compressed data,  is  capable of providing   similar performance  as  the uncompressed  detector. In \cite{Kailkhura_TSP16}, the  authors have extended  known signal detection with CS to the multiple sensor case. While intra-signal  dependence  was  considered with Gaussian measurements, the inter-sensor dependence  was neglected in \cite{Kailkhura_TSP16}. To the best of authors knowledge,  the benefits of  CS based detection when  it is difficult to perform  LR based detection  with uncompressed data due to inter-modal dependence have not been investigated in the literature.

In this paper, we seek the answer to the following question; is it  beneficial, in terms of both performance and computational complexity,  to model intermodal dependence to perform LR based detection  in the compressed domain via Gaussian approximation over either neglecting dependence (product approach) or model dependence using suboptimal methods (e.g., copula based fusion  without   knowing exactly the  best copula function that models  dependence)  with uncompressed  data?   With arbitrary marginal pdfs for each modality with uncompressed data, we show that,  under certain conditions,     better (or equivalent) detection  performance can be achieved  in the compressed domain with a small number of compressive measurements compared to performing fusion  (i). with the product approach  and (ii).   when widely available copula functions are used to model dependence of   uncompressed data.
We  briefly discuss   how to determine conditions under which  performing  compressed detection is efficient and effective  over suboptimal detection  with uncompressed data in the presence of inter-modal dependence.

\section{Detection   with Uncompressed  Data}
Let there be $L$ sensor nodes in a network deployed to solve a detection problem. The measurement vector at each node is denoted by $\mathbf x_j\in \mathbb R^N$ for $j=1,\cdots,L$.  Under hypotheses $\mathcal H_1$ and $\mathcal H_0$, $\mathbf x_j$ has the following pdfs:
%\begin{eqnarray}
%\mathcal H_1&:& \mathbf x_j = \boldsymbol \theta_j + \mathbf v_j\nonumber\\
%\mathcal H_0&:& \mathbf x_j = \mathbf v_j,  j=1,\cdots,L\label{obs_0}
%\end{eqnarray}
\begin{eqnarray}
\mathcal H_1&:& \mathbf x_j \sim f_1(\mathbf x_j)\nonumber\\
\mathcal H_0&:& \mathbf x_j \sim f_0(\mathbf x_j),  j=1,\cdots,L\label{obs_0}
\end{eqnarray}
respectively, where $f_i(\mathbf x_j)$ denotes  the joint probability density function (pdf) of $\mathbf x_j$ under $\mathcal  H_i$ for $i=0,1$ and $j=1,\cdots,L$. We assume that  the elements of $\mathbf x_j$ are independent of each other, however, the vectors $\mathbf x_j's$ are dependent for $j=1,\cdots,L$. This is a suitable model when the time samples collected at a given sensor are independent and there is spatial dependence  among sensors in a distributed network. To perform LR  based detection, it is required to compute the joint pdf of $\{\mathbf x_1, \cdots, \mathbf x_L\}$, which in general is difficult unless each $\mathbf x_j$ has a joint Gaussian pdf.

\subsection{Copula based approach}
%In \cite{}, copula theory has been exploited to estimate the joint pdf with multimodal data.
 In a parametric framework, copulas are used to construct a valid joint distribution describing an  arbitrary, possibly nonlinear dependence structure \cite{Nelsen2006}. According to copula theory,
the pdfs of $\mathbf x=\{\mathbf x_1, \cdots, \mathbf x_L\}$ under $\mathcal H_i$ can be written as \cite{Nelsen2006},
\begin{eqnarray*}
f_i(\mathbf x) = \prod_{n=1}^N \prod_{l=1}^L f_i( x_{nl}) c_{in}(u^i_{n1},\cdots,u_{nL}^i)
\end{eqnarray*}
for $i=0,1$ where $c_{in}(\cdot)$ denotes the copula density function, $u_{nl}^{i}= F(x_{nl} | \mathcal H_i)$ with $F(x|\mathcal H_i)$ denoting  the marginal cdf of $x$ under $\mathcal H_i$, and $x_{nl}$ is the $n$-th element of $\mathbf x_l$.
Then, the  log LR (LLR) can be written in the following  form:
\begin{eqnarray}
T_{LLR}(\mathbf x) &=& \log \frac{f_1(\mathbf x)}{f_0(\mathbf x)} = \sum_{l=1}^L \sum_{n=1}^N  \log \frac{f_1(\mathbf x_l[n])}{f_0(\mathbf x_l[n])} \nonumber \\
&+& \sum_{n=1}^N \log \frac{c_{1n} (u_{1n}^1, \cdots, u_{Ln}^1 | \phi_{1n})}{c_{0n} (u_{1n}^0, \cdots, u_{Ln}^0 | \phi_{0n})}\label{eq_copula_uncomp}
\end{eqnarray}
where $\phi_{1n}$ and $\phi_{0n}$ are copula parameters under $\mathcal H_1$ and $\mathcal H_0$, respectively, for $n=1,\cdots,N$. In this case, in general,  $N$ different  copulas where each one is  $L$-variate are selected to model  dependence.

One of  the fundamental challenges in copula theory is to find the  copula density function that will best fit the  given data set. Further, most of the copula density functions proposed in the literature consider the bivariate case. In order to model dependence of  multimodal data with more than two modalities, several approaches such as the use of vines  have been proposed in the literature \cite{Subramanian_2011}, which are in general computationally complex. Thus, in order to better utilize copula theory for multimodal data fusion, these challenges need to be overcome. In the following, we consider a computationally efficient approach for multimodal data fusion in which dependence among data is modeled in a low dimensional transformed domain. We  discuss  the advantages/disadvantages of the proposed approach over  the copula based approach.

%This problem has been addressed by several authors under different contexts.

\section{Detection  with  Compressed  Data}
When the signals $\mathbf x_j$'s are high dimensional, it is desired that fusion be  performed in a compressed domain. The use of low dimensional random projections for solving inference problems has been addressed in the  recent literature \cite{duarte_ICASSP06,
haupt_ICASSP07,davenport_JSTSP10,Wimalajeewa_asilomar10,
Gang_globalsip14,Bhavya_cscps14,Bhavya_asilomar14, Rao_icassp2012,Cao_Info2014,Kailkhura_WCL16,Kailkhura_TSP16}.
%Consider a reduced dimensional  sampling operator defined by a bounded linear mapping $\mathbf A_j$ of  signal $\mathbf x_j$ that lies in an ambient  Hilbert space.
Let $\mathbf A_j$  be specified by a set of unique sampling vectors  $\{\mathbf a_{j,m}\}_{m=1}^{M}$ with $M < N$ for $j=1, \cdots, L$.  Then, the low dimensional  samples can be expressed as,
\begin{eqnarray}
\mathbf y_j = \mathbf A_j \mathbf x_j   \label{obs_1}
\end{eqnarray}
  for $j=1,\cdots,L$ where $\mathbf y_j$ is the $M\times 1$ compressed  measurement vector at the $j$-th node, and the $m$-th element of the vector $\mathbf A_j\mathbf x_j$ is given by $(\mathbf A_j\mathbf x_j)_m = \langle \mathbf a_{j,m} , \mathbf x_j\rangle$ for $m=1,\cdots, M$  where $\langle., .\rangle$ denotes the inner product. In CS theory, the mapping $\mathbf A_j$ is often  selected to be a random matrix.  Solving (\ref{obs_1}) when $\mathbf x_j$'s are Gaussian is considered in \cite{davenport_JSTSP10} with a single sensor and it is extended to the multiple sensor case in \cite{Kailkhura_TSP16}.
  The degradation  of performance in the compressed domain compared to that with uncompressed data while performing LR based detection  is expressed in terms of the output SNR or the deflection coefficient in \cite{davenport_JSTSP10,Kailkhura_TSP16}. However, when $\mathbf x_j$'s are not  Gaussian and there  is dependence among them, proper performance  comparison for detection  in the  uncompressed and compressed  domains   is not available in the literature.

\subsection{LR  based detection  with compressed data}
In order to perform LR based detection   based on (\ref{obs_1}), the computation of the joint pdf of $\{\mathbf y_1, \cdots, \mathbf y_L\}$ is necessary. If the marginal pdf of $\mathbf x_j$'s are available, the marginal pdfs of each element in $\mathbf y_j$'s can be computed as in the following.  The  $m$-th element of $\mathbf y_j$, $y_{mj}$,  can be written as,
\begin{eqnarray*}
y_{mj} = \sum_{n=1}^N \mathbf A_j[m,n] x_{nj}
\end{eqnarray*}
where $\mathbf A_j[m,n]$ is the $(m,n)$-th element of $\mathbf A_j$. Having the marginal pdfs of $ x_{nj}$ and using the independence  assumption, the joint pdf of $z=  y_{mj}$ can be found after computing the characteristic function  of $z$.
It is further noted that $\{y_{mj} \}_{m=1}^M$ for a given $j$ are not necessarily  uncorrelated of each other  although $\{ x_{nj}\}_{n=1}^N$'s are uncorrelated  unless certain conditions are satisfied by $\mathbf A_j$ and $\mathbf x_j$. For example,  if the elements of $\mathbf x_j$ are  zero mean Gaussian with the covariance matrix $\sigma_v^2 \mathbf I$, and the projection matrix satisfies the condition $\mathbf A_j \mathbf A_j^T = \mathbf I$, then the elements of $\mathbf y_j$ are uncorrelated. However, in general  this uncorrelatedness   may not  hold. Once the marginal pdfs  of the elements in $\mathbf y_j$ for $j=1,\cdots,L$ are found, copula theory can be used in order to find the joint pdf of the compressive measurement vectors $\mathbf y_1, \cdots, \mathbf y_L$. Letting   $u_j = F_j( y_{qp})$ for $j=M(p-1)+q$ where  $p=1,\cdots, L$, $q=1, \cdots, M$, the LLR  based on copula functions can be expressed  as,
\begin{eqnarray}
T_{LLR}(\mathbf y)=\sum_{l=1}^L \sum_{k=1}^M   \log \frac{f_1( y_{kl})}{f_0( y_{kl})} + \log \frac{c_1 (u_1, \cdots, u_{ML} | \phi_1^*)}{c_0 (u_1, \cdots, u_{ML} | \phi_0^*)}. \label{copula_2}
\end{eqnarray}
The second term on the right hand side in  (\ref{copula_2}) requires finding  copula density functions  of $ML$ variables which is computationally very difficult. Since we assume that the elements in each $\mathbf x_j$ are independent under any given hypothesis, each element in $\mathbf y_j$ can be approximated by a Gaussian random variable (via   Lindeberg-Feller  central limit theorem assuming the required conditions are satisfied \cite{cramer_1946,thakshilaj5}) for  given $\mathbf A_j$   when $N$ is sufficiently  large. Then,  LR  based detection  can be performed via Gaussian approximation, which makes the  modeling of dependence among multimodal data with compressed measurements easier.

\subsection{LR  based detection  via Gaussian approximation}
Let $\mathbf y = [\mathbf y_1^T \cdots \mathbf y_L^T]^T$ be a $ML\times 1$ vector. With Gaussian approximation  we have  $\mathbf y|\mathcal H_i \sim \mathcal N (\boldsymbol\mu^i, \mathbf C^i)$ where
 \begin{eqnarray}
 \boldsymbol\mu^i=[{\boldsymbol\mu_1^i}^T \cdots  {\boldsymbol\mu_L^i}^T]^T
 \end{eqnarray}
 and
 \begin{eqnarray}
 \mathbf C^i =\left[
 \begin{array}{cccc}
 \mathbf C^i_1 & \mathbf C^i_{12} & \cdots & \mathbf C^i_{1L} \\
 \mathbf C^i_{21} &  \mathbf C^i_2 & \cdots & \mathbf C^i_{2L}\\
 \cdots & \cdots &  \cdots & \cdots\\
  \mathbf C^i_{L1} &  \mathbf C^i_{L2} & \cdots & \mathbf C^i_{L}
 \end{array}\right]
 \end{eqnarray}
 with $\boldsymbol\mu_j^i = \mathbb E\{\mathbf y_j | \mathcal H_i\}$,  $\mathbf C^i_j = \mathbb E\{(\mathbf y_j - \mathbb E\{\mathbf y_j\})(\mathbf y_j - \mathbb E\{\mathbf y_j\})^T | \mathcal H_i\}$, $\mathbf C^i_{jk} = \mathbb E\{(\mathbf y_j - \mathbb E\{\mathbf y_j\})(\mathbf y_k - \mathbb E\{\mathbf y_k\})^T | \mathcal H_i\}$ with  $j\neq k$, $k=1,\cdots,L$ and $j=1,\cdots,L$ for $i=0,1$.  Further, let $\boldsymbol\beta_j^i = \mathbb E\{\mathbf x_j | \mathcal H_i\}$, $\mathbf D^i_j = \mathbb E\{(\mathbf x_j - \mathbb E\{\mathbf x_j\})(\mathbf x_j - \mathbb E\{\mathbf x_j\})^T | \mathcal H_i\}$ and $\mathbf D^i_{jk} = \mathbb E\{(\mathbf x_j - \mathbb E\{\mathbf x_j\})(\mathbf x_k - \mathbb E\{\mathbf x_k\})^T | \mathcal H_i\}$  for   $j\neq k$. Then we  have,
 \begin{eqnarray}
 \boldsymbol\mu_j^i &=& \mathbf A_j \boldsymbol\beta_j^i,
 \mathbf C^i_j = \mathbf A_j \mathbf D^i_j \mathbf A_j^T, \mathrm{and~}
 \mathbf C^i_{jk} = \mathbf A_j \mathbf D^i_{jk} \mathbf A_k^T
 \end{eqnarray}
for $j,k=1,\cdots, L$ and $i=0,1$. Then, we can write,
\begin{eqnarray}
\boldsymbol\mu^i = \mathbf A \boldsymbol \beta^i \mathrm{~and~}
\mathbf C^i = \mathbf A \mathbf D^i \mathbf A^T
\end{eqnarray}
where
\begin{eqnarray}
\mathbf A = \left(
\begin{array}{ccccc}
\mathbf A_1 & \mathbf 0 & \cdot  & \cdot &\mathbf 0\\
\mathbf 0 & \mathbf A_2 & \cdot  & \cdot &\mathbf 0\\
 \cdot  &  \cdot  & \cdot  & \cdot & \cdot \\
 \mathbf 0 & \mathbf 0 & \cdot  & \cdot & \mathbf A_L\\
%\mathbf A_1, & \mathbf 0 &\dot &\mathbf 0
\end{array}\right)
\end{eqnarray}
is a $ML \times NL$ matrix and $\boldsymbol\beta^i$ and $\mathbf D^i$ are notations analogous  to $\boldsymbol\mu^i$ and $\mathbf C^i$, respectively. Then, the decision statistic of the LLR based detector is simply  given by,
\begin{eqnarray*}
\Lambda = \mathbf y^T ({\mathbf C^1}^{-1} - {\mathbf C^0}^{-1})\mathbf y - 2 ({\boldsymbol\mu^1}^T {\mathbf C^1}^{-1} - {\boldsymbol\mu^0}^T {\mathbf C^0}^{-1}) \mathbf y.
\end{eqnarray*}
To illustrate  the detection performance with multimodal data  in the compressed domain with Gaussian approximation compared to detection  with uncompressed data, in the following, we present   a numerical example  considering $L=2$. We further consider the  elements of $\mathbf A_j$ to be iid zero mean Gaussian for $j=1,2$.

\subsection{Example}\label{example1}
We consider two cases. In Case I, $\mathbf x_1$, and  $\mathbf x_2$  have the following  marginal pdfs  under the two hypotheses (as considered  in \cite{iyengar_tsp11}): $x_{i1} | \mathcal H_j \sim \mathcal N(0,\sigma_j^2)$, and  $x_{i2} | \mathcal H_j \sim \mathrm{Exp} (\lambda_j)$. It is noted that   $x\sim \mathrm{Exp} (\lambda)$ denotes that $x$ has an  exponential distribution with $f(x) = \lambda e^{-\lambda x}$ for $x\geq 0$ and $0$ otherwise.  Under  $\mathcal H_1$, $x_{i2}$'s are generated so that
$
x_{i2} =x_{i1}^2+w^2
$
where $w\sim \mathcal N(0, \sigma_1^2)$. Then we have $x_{i2}\sim \mathrm{Exp} (\lambda_1)$ with $\lambda_1=\frac{1}{2\sigma_1^2}$. Under $\mathcal H_0$,  $x_{i2}$'s are generated independent of $x_{i1}$ for $i=1,\cdots,N$ with parameter $\lambda_0$.

For Case II, we consider that $x_{i1} \sim \mathrm{Exp} (\lambda_j)$  and $x_{i2} | \mathcal H_j \sim \mathrm{Beta}(a_i, b_i=1)$ where $x\sim \mathrm{Beta}(a, b)$ denotes that $x$ has a beta distribution with pdf $f(x) = \frac{1}{\mathcal B(a,b)}x^{a-1} (1-x)^{b-1}$ and  $\mathcal B(a,b) = \frac{\Gamma(a)\Gamma(b)}{\Gamma(a+b)}$ is the beta function. Under $\mathcal H_1$, $x_{i2} $'s are  generated so that \begin{eqnarray*}
x_{i2} =\frac{u}{u+x_{i1}}
\end{eqnarray*}
where $u\sim \mathrm{Gamma}(\alpha_1, \beta_1=1/\lambda_1)$.  Then $x_{i2} | \mathcal H_1 \sim \mathrm{Beta}(a_1, b_1=1)$ with $a_1=\alpha_1$. It is noted that $x\sim \mathrm{Gamma}(\alpha, \beta)$ denotes that $x$ has Gamma pdf with $f(x) = \frac{1}{\beta^{\alpha}\Gamma(\alpha)} x^{\alpha-1} e^{-x / \beta}$ for $x\geq 0$ and $\alpha, \beta > 0$. Under $\mathcal H_0$, $x_{i2}$ is generated independent of $x_{i1}$ with parameters $a_0$ and $b_0=1$.

\begin{figure}
  \centering
 \includegraphics[width = 3.25in, height=!]{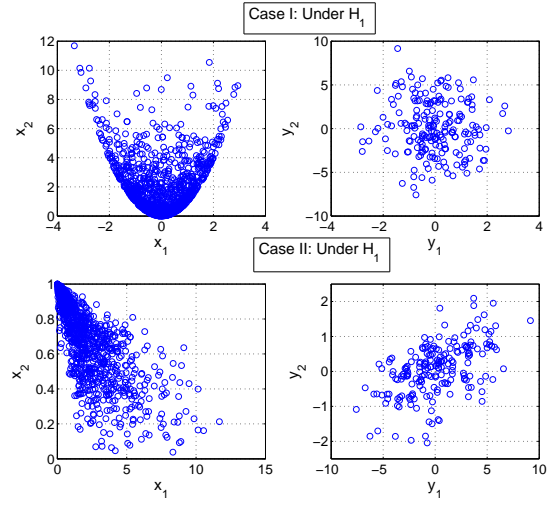}
\caption{Scatter plots of uncompressed and compressed data under $\mathcal H_1$; $N=1000$, $M=200$, $L=2$}\label{fig:scatter_L2}
\end{figure}

First,  we illustrate  how the dependence  structure of the  data  changes from uncompressed domain to the compressed domain. In  Fig.  \ref{fig:scatter_L2},  we show the scatter plots for both compressed and uncompressed data at the  two sensors under $\mathcal H_1$.  In Fig.  \ref{fig:scatter_L2}, the top and bottom subplots are for Case I and Case II, respectively while left and right subplots are for uncompressed and compressed data, respectively. It can be observed that while uncompressed data at the two sensors are strongly dependent of each other,  compressed data appears to be  weakly dependent with a completely different (Gaussian like)  pattern.

\begin{figure}[h!]
    \centering
    \begin{subfigure}[b]{0.35\textwidth}
        \includegraphics[width=\textwidth]{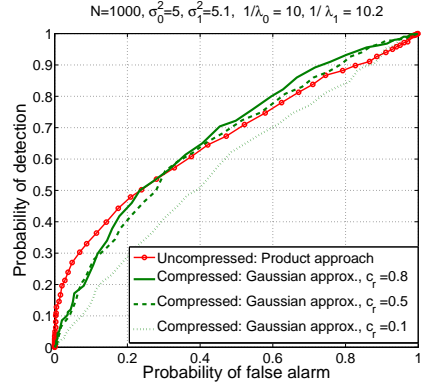}
        \caption{ Case I: $\sigma_0^2=5$, $\sigma_1^2=5.1$, $1/\lambda_0=10$, $1/\lambda_1=10.2$}
        \label{fig:N100}
    \end{subfigure}
%\hfil
    ~ %add desired spacing between images, e. g. ~, \quad, \qquad, \hfill etc.
      %(or a blank line to force the subfigure onto a new line)
    \begin{subfigure}[b]{0.35\textwidth}
        \includegraphics[width=\textwidth]{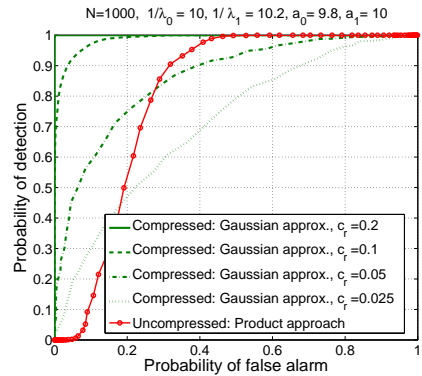}
        \caption{ Case II: $1/\lambda_0=10$, $1/\lambda_1=10.2$, $a_0=9.8$, $a_1=10$}
        \label{fig:N1000}
    \end{subfigure}

    \caption{Detection performance with   multimodal dependent data in the compressed and uncompressed domains:  $N=1000$}\label{fig:ROC_case_I_II}
\end{figure}

\subsection{Product approach with uncompressed data vs. Gaussian approximation with compressed data}
In the following, we compare the detection  performance with  compressed multimodal data  and  the product approach (where   dependence  is ignored) with uncompressed data. Fig. \ref{fig:ROC_case_I_II} shows the  performance in terms of the ROC curves for the two cases considered in Example \ref{example1}. We make several important observations here.  In Case I, the detection performance with  the Gaussian approximation in the uncompressed domain is only slightly better than that with the product approach in the uncompressed domain when the compression ratio, $c_r = \frac{M}{N}$,  is relatively large and the probability of false alarm is high. For small $c_r$, the product approach with uncompressed data shows  better performance  than the Gaussian approximation, however, the performance gap is not very significant.  In Case II, we observe a significant performance gain when performing detection  with compressed data even with relatively small $c_r$ compared to the product approach in the uncompressed domain. It is noted that in Case I, the observations at the two sensors are uncorrelated with uncompressed data (although they are dependent) thus $\mathbf D^1$ is diagonal. Thus, not taking  dependence into account  in the uncompressed domain seems not to result in a large performance loss compared to taking dependence in  the compressed domain into account. On the  other hand, when considering Case II, it is noted that  $\mathbf D^1$ is not diagonal, and the uncompressed observations under $\mathcal H_1$ are strongly  correlated. Thus, ignoring dependence with uncompressed data leads to severe performance loss compared to taking dependence (via Gaussian approximation) into account in the compressed domain even with very small $c_r$. Further, in that case, it is observed that, there is a threshold for  $c_r$ after which the Gaussian approximation in the compressed domain starts to perform better than the product approach with uncompressed data.

\begin{figure}
\centerline{\epsfig{figure=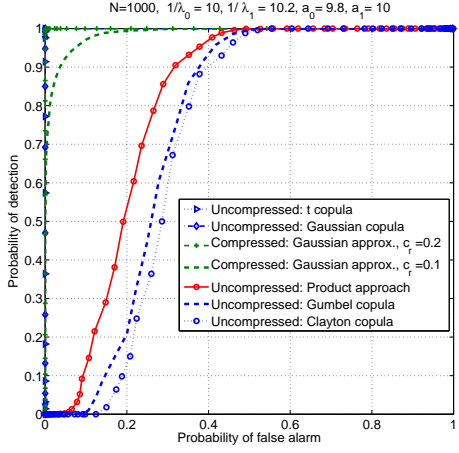,width=7.0cm}}
\caption{Detection performance with   multimodal dependent data in the compressed and uncompressed domains for case II; $1/\lambda_0=10$, $1/\lambda_1=10.2$, $a_0=9.8$, $a_1=10$, $b_0=b_1=1$ }\label{fig_copula}
\end{figure}
\subsection{Copula based fusion  with uncompressed data vs. Gaussian approximation with compressed data}
Next, we compare the detection  performance  when copulas are used to compute the joint pdf  with uncompressed data in Fig. \ref{fig_copula}. Since finding optimal copula function that models a  given data  set is computationally complex, we plot the detection performance using  widely available bivariate copula functions. To that end, we consider Gaussian, t, Gumbel and Clayton copula functions  as described in \cite{iyengar_tsp11,He_tsp2015}. Further, we consider Example \ref{example1} with Case II. We further plot the detection  performance with the product approach with uncompressed data. It is observed from Fig. \ref{fig_copula} that fusion with Gaussian and t copula functions leads to perfect detection, while fusion with  Gumbel and Clayton copula provides poor performance even compared to the product approach. On the other  hand, fusion performance with compressed data with $c_r=0.2$ is capable of providing perfect detection with the  parameters considered. Thus, with the  considered problem parameters, the use of copula functions  with uncompressed data seems to be a waste of resources when  perfect detection can be achieved with less computational complexity in the compressed domain via Gaussian approximation. Thus, it is worthwhile to investigate as to  when it is beneficial to use copula theory to model dependence with uncompressed data compared to performing  fusion by  modeling dependence  with compressed data in a computationally easier fashion. We briefly address this issue in the  following.

%\subsection{Performance comparison: uncompressed data vs. Gaussian approximation in the  compressed domain}
In order to quantify the performance of detection with both uncompressed and compressed data, we consider Kullback-Leibler (KL)  distance to be the performance metric. The KL distance between the pdfs under the two hypotheses in the compressed domain  with the  Gaussian approximation  can be computed as  \cite{cover1}
 \begin{eqnarray}
&~& \mathcal D_{KL}^{c,G} (f_0 || f_1)\nonumber\\
% &=& \frac{1}{2}\left\{\mathrm{tr}({\mathbf C^1}^{-1}\mathbf C^0) + (\boldsymbol\mu^1 -\boldsymbol\mu^0 )^T {\mathbf C^1}^{-1} (\boldsymbol\mu^1 -\boldsymbol\mu^0 ) \right.\nonumber\\
% &~&\left.- ML  + \log \frac{|\mathbf C^1|}{|\mathbf C^0|} \right\}\nonumber\\
  &=& \frac{1}{2}\left\{\mathrm{tr}(\mathbf A^{\ddag}\mathbf D^0) + (\boldsymbol\beta^1 -\boldsymbol\beta^0 )^T \mathbf A^{\ddag} (\boldsymbol\beta^1 -\boldsymbol\beta^0 ) \right.\nonumber\\
 &~&\left.- ML  + \log \frac{|\mathbf A\mathbf D^1 \mathbf A^T|}{|\mathbf A\mathbf D^0 \mathbf A^T|} \right\}
 \end{eqnarray}
where $\mathbf A^{\ddag} = \mathbf A^T (\mathbf A \mathbf D^1 \mathbf A^T)^{-1} \mathbf A$ and $\mathrm{tr}(\cdot)$ denotes the trace operator. In the case where   $\mathbf x_l$'s for $l=1,\cdots, L$ are assumed to be independent of each other  under $\mathcal H_0$, we have $f_0(\mathbf x) = \underset{l,n}{\prod} f_0^m(x_{nl})$ where $f_i^m$ denotes the marginal pdf under $\mathcal H_i$. Thus, the KL distance between $f_0(\cdot)$ and $f_1(\cdot)$ with uncompressed data can be written as,
\begin{eqnarray*}
 \mathcal D_{KL}^u (f_0 || f_1) &=&  \mathcal D_{KL}^{u,p} (f_0 || f_1^m) \nonumber\\
 &-& \underbrace{\mathbb E \left\{\sum_{n=1}^N \log c_{1n} (u_{1n}^1, \cdots, u_{Ln}^1 | \phi_{1n})| \mathcal H_0\right \}}_{\Upsilon_{f_0,c} }
\end{eqnarray*}
where  $ \mathcal D_{KL}^{u,p} (f_0 || f_1^m)$ denotes the KL distance under the product approach.
%It is noted that,
%\begin{eqnarray}&~&
%\mathbb E \left\{\sum_{n=1}^N \log c_{1n} (u_{1n}^1, \cdots, u_{Ln}^1 | \phi_{1n}) | \mathcal H_0 \right \} \\ &=&\sum_{n=1}^N  \mathbb E \left\{\log c_{1n} (u_{1n}^1, \cdots, u_{Ln}^1 | \phi_{1n}) | \mathcal H_0\right\}
%\end{eqnarray}
When the  marginal pdfs are available, $ \mathcal D_{KL}^{u,p} (f_0 || f_1^m) $ can be computed.  It is noted that the term $ \Upsilon_{f_0,c}$ depends on the particular copula function used to model  dependence.
Thus, for  a given copula function, when
$
\Upsilon_{f_0,c} > \mathcal D_{KL}^{u,p} -\mathcal D_{KL}^{c,G}
$ performing detection in the compressed domain with given $M$ ($\mathcal D_{KL}^{c,G}$ is a function of $M$) appears to be  more effective and efficient than copula based fusion  in the uncompressed domain. This issue will be further addressed  in detail in future work.

\section{Conclusion}
In this paper, we showed that, under certain conditions,   detection with multimodal dependent data with compressive sensing  can be  better (or equivalent)  than  detection with the widely considered product approach and  copula based fusion    with uncompressed data. We briefly discussed the conditions under which   modeling dependence for likelihood ratio based detection   in the compressed domain is more efficient and effective than modeling dependence with uncompressed data using copula theory  which is computationally expensive  most of the time. Experiments with real datasets will be considered in future work.

\newpage

\bibliographystyle{IEEEtran}
\bibliography{bib1,ref_1}
%\bibliography{ref_1}

\end{document}